\documentclass[12pt, 3p, review sort&compress]{elsarticle}

\usepackage{amsmath}
\usepackage{xspace}\newcommand{\etal}{\emph{et al.}\xspace}
\usepackage{todonotes}
\usepackage{soul}
\usepackage{subcaption}
\usepackage{xcolor}
\usepackage{amsmath, amssymb}
\usepackage{booktabs}
\usepackage[normalem]{ulem}
\usepackage{hyperref}

\newcommand{\stkout}[1]{\ifmmode\text{\sout{\ensuremath{#1}}}\else\sout{#1}\fi}

\newcommand{\uu}[1]{\ensuremath{\,\mathrm{#1}}}

\begin{document}

\begin{frontmatter}

\journal{Computational Materials Science}

\title{Amending CALPHAD databases using a neural network for predicting mixing enthalpy of liquids}

\author[cdl]{Clement Vincely\corref{cor}}
\ead{clement.vincely@unileoben.ac.at}
\cortext[cor]{Corresponding author}
\author[mul]{Amin Reiners-Sakic}
\author[cps]{Vedant Dave}
\author[cdl-tuw]{Erwin Povoden-Karadeniz}
\author[cps]{Elmar Rueckert}
\author[cdl]{Ronald Schnitzer}
\author[cdl]{David Holec}

\affiliation[cdl]{organization={Christian Doppler Laboratory for Knowledge-based Design of Advanced Steels (CDL-KnowDAS), Department of Materials Science, Montanunivestität Leoben}, addressline={Franz-Josef-Straße 18}, city={Leoben}, postcode={8700}, country={Austria}}

\affiliation[mul]{organization={Department of Materials Science, Montanunivestität Leoben}, addressline={Franz-Josef-Straße 18}, city={Leoben}, postcode={8700}, country={Austria}}

\affiliation[cps]{organization={Cyber Physical Systems Lab, Montanunivestität Leoben}, addressline={Franz-Josef-Straße 18}, city={Leoben}, postcode={8700}, country={Austria}}

\affiliation[cdl-tuw]{organization={Christian Doppler Laboratory for Interfaces and Precipitation Engineering (CDL-IPE), Institute of Materials Science and Technology, TU Wien}, addressline={Getreidemarkt 9}, city={Vienna}, postcode={1060}, country={Austria}}

\begin{abstract}
In order to establish the thermodynamic stability of a system, knowledge of its Gibbs free energy is essential. 
Most often, the Gibbs free energy is predicted within the CALPHAD framework using models employing thermodynamic properties, such as the mixing enthalpy, heat capacity, and activity coefficients. 
Here, we present a deep-learning approach capable of predicting the mixing enthalpy of liquid phases of binary systems that were not present in the training dataset.
Therefore, our model allows for a system-informed enhancement of the thermodynamic description to unknown binary systems based on information present in the available thermodynamic assessment. 
Thereby, significant experimental efforts in assessing new systems can be spared.
We use an open database for steels containing 91 binary systems to generate our initial training (and validation) and amend it with several direct experimental reports.
The model is thoroughly tested using different strategies, including a test of its predictive capabilities.
The model shows excellent predictive capabilities outside of the training dataset as soon as some data containing species of the predicted system is included in the training dataset.
The estimated uncertainty of the model is below $1\uu{kJ/mol}$ for the predicted mixing enthalpy.
Subsequently, we used our model to predict the enthalpy of mixing of all binary systems not present in the original database and extracted the Redlich-Kister parameters, which can be readily reintegrated into the thermodynamic database file.
\end{abstract}




\end{frontmatter}

\section{Introduction}
\label{sec:introduction}

Phase diagrams provide crucial information on phases, thermodynamic regions of phase stability, and phase transitions, which is essential for materials development and production. 
The most widely used approach for phase diagram simulations is the CALPHAD (CALculation of PHase Diagrams) method, which uses a parameterized form of the Gibbs energy stored in thermodynamic databases~\cite{lukas2007CALPHAD}.
According to the CALPHAD method, the total Gibbs free energy, $G^{\text{tot}}$, of a system is the sum of the elemental Gibbs free energies, $G^{\text{ref}}=\sum_{i}{}^0G^{\text{ref}}_i$), the configurational entropy term $-TS^{\text{id}}$ (corresponding to $G^{\text{id}}$, Gibbs free energy of mixing in \textit{ideal} solution with $H^{\text{id}}=0$), and the excess Gibbs free energy, $G^{\text{excess}}$. 
The latter two terms describe the non-interacting and interacting parts of the mixing Gibbs energy~\cite{lukas2007CALPHAD}.

While $G^{\text{ref}}$ and $S^{\text{id}}$ are readily accessible, $G^{\text{excess}}$ requires fitting to experimental and/or theoretical data. 
Obtaining thermodynamic data, such as mixing enthalpy, $H_{\mathrm{mix}}$, and heat capacity, for assessment purposes is both time-consuming and expensive. 
Furthermore, while the existing literature provides extensive data for some systems, such as Fe-Ni~\cite{https://doi.org/10.1002/srin.199801599} and Al-Mg~\cite{library978286}, many other systems remain sparsely described. 

One tempting approach is to use machine learning (ML) techniques.
Recently, Laiu~\etal~\cite{Laiu2022} developed an ML approach using neural networks (NNs) to predict $G^{\text{tot}}$ for ternary face-centered cubic solid solutions based on binary $G^{\text{tot}}$ data. 
A similar approach was also recently applied to liquid phases using different ML approaches, including a gradient boosting decision tree, NN, and a Gaussian process to predict the $H_{\mathrm{mix}}$ of binary systems and also to extend it to ternaries. 
Both works demonstrate that ML approaches can be used to significantly speed up the thermodynamic description of multicomponent systems and also to extrapolate to higher-order systems~\cite{DEFFRENNES2024102745}. 
However, in the triangular compositional space, the above works mean \emph{interpolating} the interior of the triangle from the known behavior on all three sides, i.e., binary systems. CALPHAD bases the extended system information on the physics of unaries and binaries.
Often, however, certain binary systems, which may not have been important in the past but will become more relevant in the future, are not yet described in thermodynamic databases. 
An example of this is systems containing tramp elements, such as Fe-Sn, Fe-Sb, and Fe-As, which are becoming more and more important for research~\cite{sakic2024interplay} as well as for steel producers~\cite{dworak2023stahlrecycling} due to the increasing amount of scrap recycling in steel production.

To circumvent the thermodynamic description of these systems from extensive experimental and theoretical studies, the aim of this work is to develop a ML model that is able to predict the unknown term $G^{\text{excess}}$ along one of the triangle sides as an \emph{extrapolation} from the other two known sides. 
More precisely, we focus on predicting the enthalpy part $H^{\text{excess}}=H^{\text{mix}}$ using an NN, even for systems not directly covered by the training data. 
Importantly, we also present a method for quantification of the uncertainty in our predictions. 
We then fit the predicted $H_{\text{mix}}$ directly to a Redlich-Kister (RK) polynomial, as is common in the CALPHAD community. 
By adding $G^{\text{ref}}$ and $G^{\text{id}}$, it is finally possible to obtain a complete parameterization of $G^{\text{tot}}$, which can then be reintegrated into an existing thermodynamic database.  

Previous works suggested that the liquid phase often provides more accurate and easier accessible experimental thermodynamic mixing enthalpies than solid phases~\cite{dumitrescu2016CALPHAD, koch1992database}.
Therefore, in this work, we focus on the liquid phase and use it as a proof of concept for our ML model being able to extrapolate outside of the training dataset.
We utilize a freely available steel database provided by \texttt{MatCalc} Engineering~\cite{matcalc} as a training dataset as well as for validation purposes.
This database contains data for 91 binary liquid phases containing 22 elements. 
In contrast to other works~\cite{Laiu2022, DEFFRENNES2024102745}, we also amend the training dataset with experimental $H_{\mathrm{mix}}$ values from the literature. 
This reduces the danger of training the model with wrong or inaccurate data stemming from the used thermodynamic database.
After training and validating our model, we provide predictions for the complete matrix of 231 binary liquids.

\section{Methodology}
\label{sec:methods}

\subsection{Preprocessing: Data collection and feature extraction}
\label{sec:data_preprocessing}

\begin{figure}[h!]
    \centering
    \includegraphics[width=0.5\textwidth]{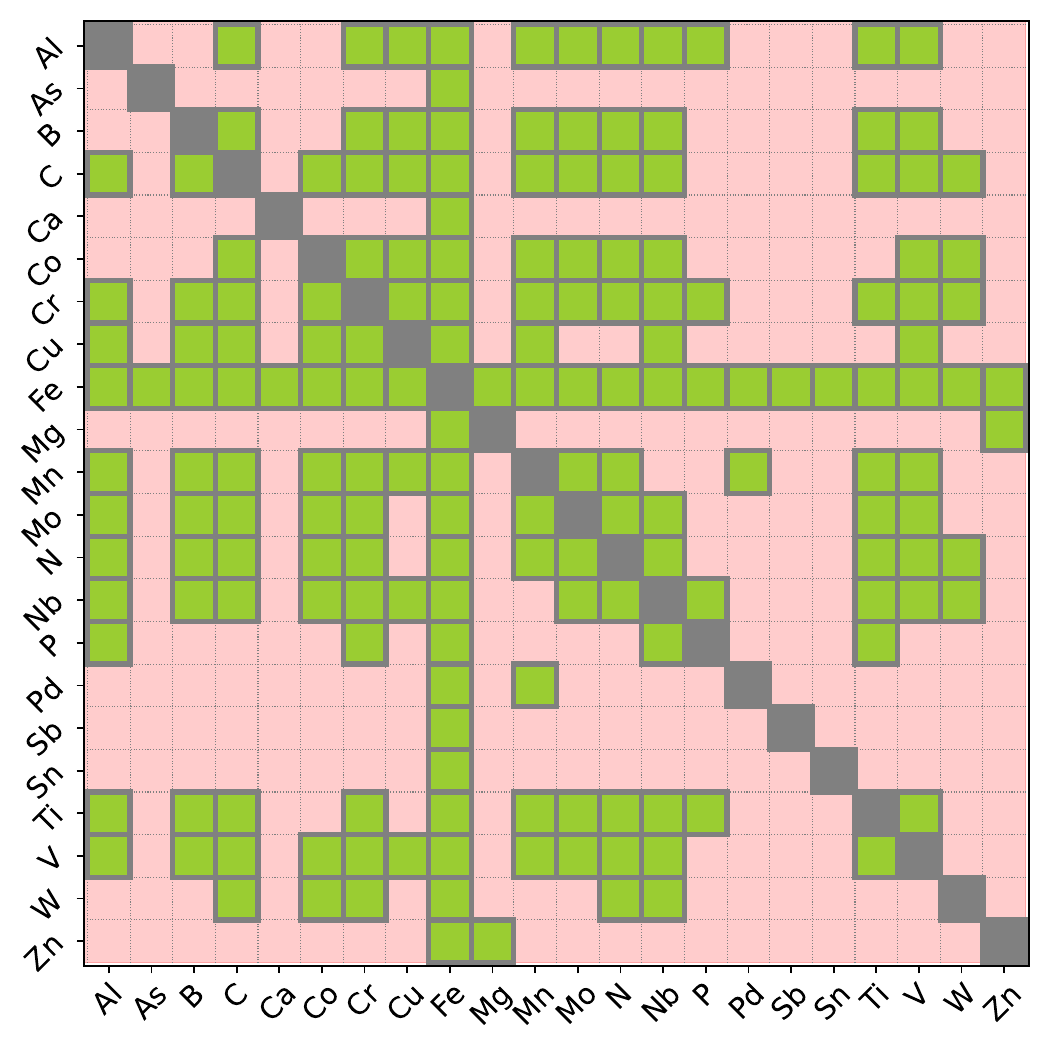}
    \caption{This heatmap provides information on the liquid binary phases in the used, open-source thermodynamic database file~\cite{matcalc}. 
    The green and red fields indicate systems for which a thermodynamic description exists or is missing.}    
    \label{fig:systems}
\end{figure}

We utilized the \texttt{MatCalc} database for steel with trace and tramp elements~\cite{matcalc} to extract $H_{\mathrm{mix}}$ data at fixed temperature for 91 binary systems (Fig.~\ref{fig:systems}), which formed the basis of our training and validation dataset.
This step was done using the \texttt{pycalphad} library~\cite{otis2017pycalphad}.
2000 compositional equidistant $H_{\mathrm{mix}}$ data points were extracted for every binary liquid phase in the thermodynamic database.
We chose $T=1873\uu{K}$ for this work, which is a commonly used reference temperature in thermodynamic assessments~\cite{Dreval2015}.
Additional experimental $H_{\mathrm{mix}}$ data were collected for various liquid binary systems from \cite{TanakaGokcenMorita+1990+49+54, Turchanin2003, 1981925}.
We therefore do not rely on the thermodynamic database as the only data source, but also supplement the data set with literature values.
All collected data are tabulated and provided through the project's online data repository~\cite{data_share}.

Next, we utilized \texttt{Matminer}~\cite{ward2018matminer} to extract a comprehensive set of 151 \textit{atomic-based features}.
Among others, these include properties such as mean electronegativity, ground-state bandgap, and atomic mass.
Previous studies have demonstrated the effectiveness of using such features for predicting various material properties, including compound $H_{\mathrm{mix}}$~\cite{jha2018elemnet}.

Additionally, 181 \textit{composition-based features} derived from the \texttt{Matminer} library are employed. 
In this context, we highlight the \texttt{ValenceOrbital} and \texttt{AtomicOrbitals} features that describe the electronic configuration and orbital overlaps that underlie bond formation tendencies. 
Additionally, \texttt{ElectronegativityDiff} quantifies disparities in elemental affinities, which often correlate with enthalpic stability or repulsion in the liquid state. 
With these features, we expect our ML model to be able to identify complex nonlinear relationships between composition and $H_{\mathrm{mix}}$.
Furthermore, we also included Miedema and Yang-based alloy features in the composition-based descriptor set via \texttt{Matminer}.

The Miedema features directly encode estimations of chemical, elastic, and structural contributions to $H_{\mathrm{mix}}$, making them particularly relevant for capturing alloy thermodynamics in both solid and liquid phases. 
Complementarily, Yang’s descriptors (\texttt{YangSolidSolution} featurizer) focus on the solubility and phase stability in complex alloys.
This dual approach of incorporating both atomic-based and composition-based features is supported by recent advancements in materials informatics~\cite{liu2017materials}.

For preprocessing the features, we used the \texttt{StandardScaler} from the \texttt{scikit-learn} package~\cite{scikit-learn}.

\subsection{Neural network geometry}
\label{sec:hyperparameters}

\begin{figure}[h!]
    \centering
    \includegraphics[width=\textwidth]{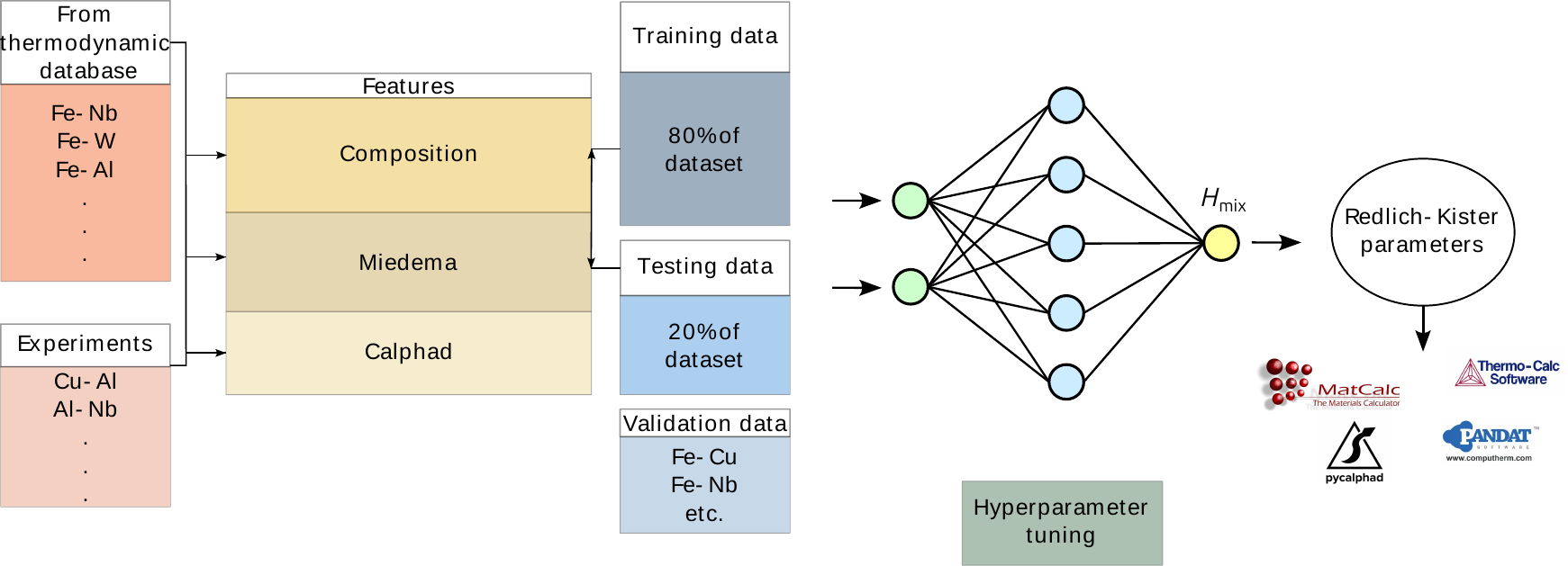}
    \caption{The workflow representing the featurization and validation process to construct a NN for extracting the mixing enthalpy ($H_{\mathrm{mix}}$), which is subsequently fitted by the Redlich-Kister (RK) polynomial.}
    \label{fig_nn}
\end{figure}

The implementation of the NN (see Fig.~\ref{fig_nn}) and hyperparameter optimization was carried out using Python-based ML libraries. 
Specifically, the model was built using \texttt{TensorFlow}~\cite{tensorflow2015-whitepaper}. 
The high-level API \texttt{Keras}~\cite{chollet2015keras}, integrated within \texttt{TensorFlow}, was utilized to streamline the model architecture definition and training procedures. 
Hyperparameters, such as the number of layers, the number of neurons per layer, learning rates, and activation functions, determine the model architecture.
To optimize them, we utilized \texttt{KerasTuner}~\cite{omalley2019kerastuner}, which allows for their automated and systematic exploration employing a random search algorithm. 
The parameter space included configurations with one to ten layers and one to 512 nodes per layer~\cite{bergstra2012random}. 
The optimal hyperparameters were selected based on their ability to minimize the loss function and improve predictive accuracy on validation data, thereby preventing overfitting of the model.
The resulting parameters defining the architecture of our NN are given in Supplementary Material Table~S1.

\subsection{Neural network validation}
\label{sec:NN_model}

We partitioned the dataset into training and validation sets to ensure the robustness and generalizability of our NN model.
To do so, we used the standard $k$-fold cross-validation test to assess the model's performance metrics (see Sec.~\ref{sec:k-fold})

In addition to randomly selecting data points for the validation set, we also performed leave-one-out cross-validation (LOOCV) (see Sec.~\ref{sec:LOOCV}).
To do so, we quarantined entire binary systems, hence the model is evaluated on binary systems it has never encountered during training, thus providing a stringent test of its predictive capabilities~\cite{stone1974cross}. 
This method has proven effective in assessing model performance on truly unseen data, particularly in materials science applications~\cite{ward2016general}.

\subsection{Redlich-Kister parameterization}
\label{sec:redlich-kister}

Since our primary aim is to provide a tool for extending existing thermodynamic databases, we use the RK polynomial to fit the predicted $H_{\mathrm{mix}}$ values obtained from the NN model.
Thereby, we mimic the common practice in CALPHAD assessments, where a polynomial is fitted to experimental data.

The general form of the RK polynomial approximating $H_{\mathrm{mix}}$, of a binary system is given by:
\begin{equation}
    H_\text{mix} = x_1 x_2 \sum_{k=1}^{n} L_k (x_1 - x_2)^{(k-1)}\ ,
    \label{eq:redlich-kister}
\end{equation}
where \( x_1 \) and \( x_2 \) are the mole fractions of the two components ($x_1+x_2=1$), \( L_k \) are the RK parameters, and \( n \) is the order of the polynomial; in the present work we use $n=4$ uniformly for all studies systems. 
All obtained RK parameters are listed in Supplementary Material Table S3.

\section{Results and discussion}
\label{sec:results}

\subsection{$K$-fold cross-validation}
\label{sec:k-fold}
In order to validate our NN on the entire training dataset, we performed a 10-fold cross-validation. 
The dataset was randomly split into 10 equal-sized subsets, where the model was trained on nine subsets and tested on the remaining one.
The test was performed 10 times, rotating through all subsets. 
A comparison of the obtained models is provided in Supplementary Table~S2. 

\begin{figure}[h!]
    \centering
    \includegraphics[width=0.8\linewidth]{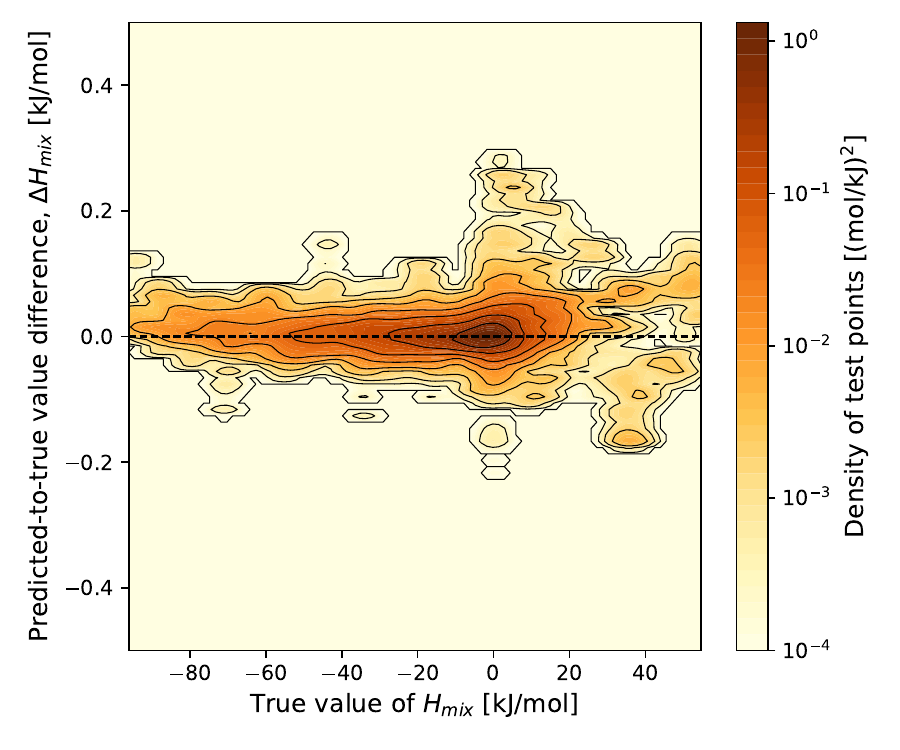}
    \caption{Overview of the best-performing $k$-fold model (Fold 7) shown as the density of points representing the difference between predicted and true value ($\Delta H_{\mathrm{mix}}$, $y$-axis) vs. the true value $H_{\mathrm{mix}}$ ($x$-axis).}
    \label{fig:parity_plot}
\end{figure}

The best model in terms of performance is Fold 7, which achieves an exceptionally low Mean Absolute Error (MAE) of $16.52\uu{J/mol}$, Mean Squared Error (MSE) of $668.14\uu{(J/mol})^2$, and Root Mean Squared Error (RMSE) of $25.85\uu{J/mol}$. 
These results suggest the model's excellent predictive accuracy with minimal error. 
Additionally, the consistency of the performance metrics across all folds indicates that our NN is not overfitting. 
For instance, the other folds, including those with the maximum values of MAE (Fold 9 with MAE of $88.90\uu{J/mol}$, and Fold 10 with MAE of $50.50\uu{J/mol}$), show acceptable results without large deviations.
The absence of significant variation in performance between folds suggests that the model is stable and reliable across different subsets of the data, demonstrating no signs of overfitting.
The Fold 7 is the model that we use for the final predictions presented in Sec.~\ref{sec:redlich-kister_results}.
It's performance is further visualized in Fig.~\ref{fig:parity_plot} showing that the discrepancies between true and predicted values, $\Delta H_{\text{mix}}$, are consistently below $0.2\uu{kJ/mol}$ for a wide range of $H_{\text{mix}}$ between $-100\uu{kJ/mol}$ and $50\uu{kJ/mol}$. This high level of agreement between predicted and actual values on the unseen testing set underscores the model’s robustness and generalization capacity.

\subsection{LOOCV validation of the model}
\label{sec:LOOCV}
As a first test of the predictive power of our model when extrapolating to new binary systems, we employed the LOOCV procedure. 
Here we used the same NN architecture as before, but this time trained the model from scratch with one binary system completely excluded from the training set. 
The thus quarantined data were subsequently used for validation purposes.

\begin{figure}[h!]
    \centering
    \includegraphics[width=0.8\textwidth]{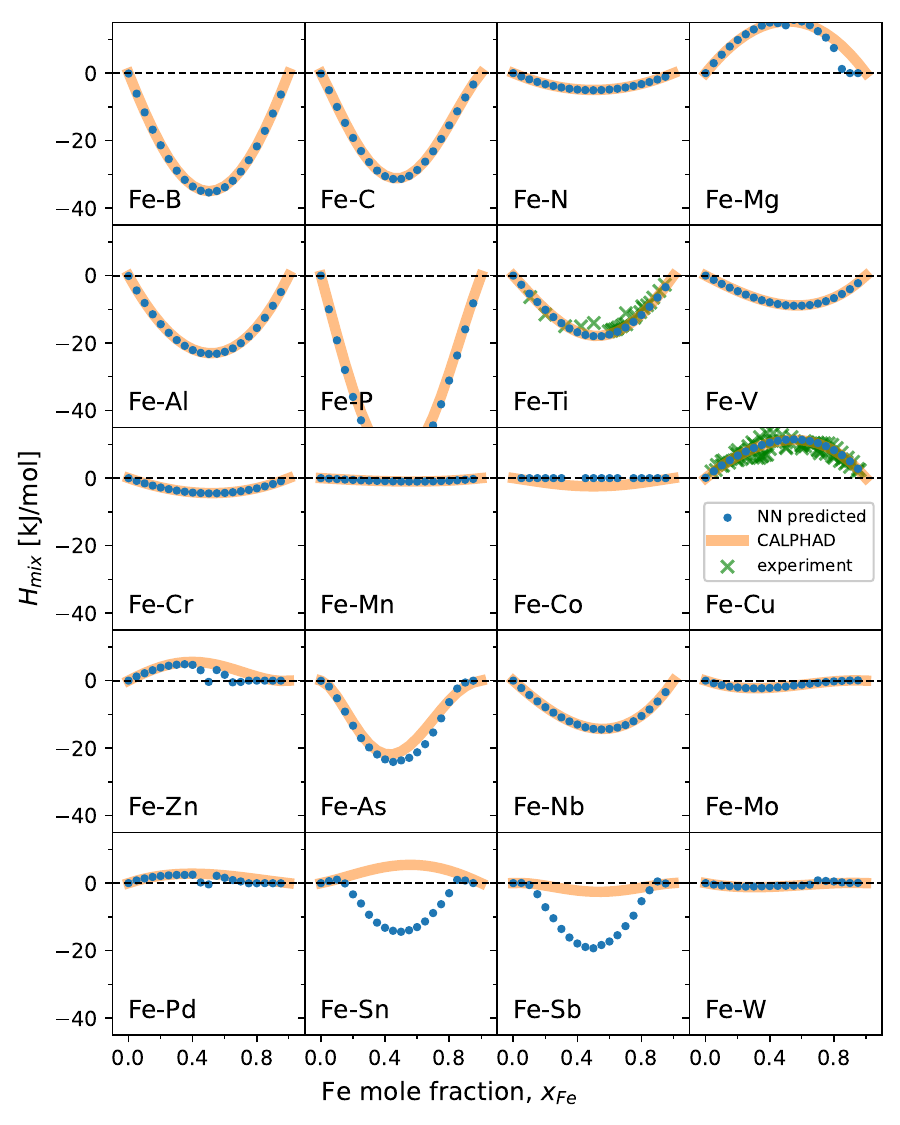}
    \caption{LOOCV predictions of the mixing enthalpy ($H_{\mathrm{mix}}$) for binary Fe-$X$ systems in liquid phase at 1873\,K exemplified for most of the Fe-$X$ systems.
    The blue points present the actual predictions with the individual NN (trained without the Fe-$X$ data), the orange curve and green crosses show the source CALPHAD and collected experimental data, respectively.
    }
    \label{fig:Fe-X_LOOCV}
\end{figure}

For example, all the CALPHAD and experimental data points corresponding to the binary system of Fe-Cu were quarantined during the training process, and the resultant NN prediction was then compared against those values (see right column, third row in Fig.~\ref{fig:Fe-X_LOOCV}).
The NN-predicted $H_{\mathrm{mix}}$ shows a good match with the CALPHAD values (orange line).
Not only does it reproduce qualitatively the demixing of the Fe-Cu system, but it also qualitatively exhibits excellent agreement, particularly considering the spread of the experimental data (green crosses); in this case, it is even larger than a typical errorbar of $\approx2\,\text{kJ/mol}$.
Our model qualitatively as well as quantitatively predicts all one-by-one quarantined systems, with Fe-Sn and Fe-Sb (bottom row in Fig.~\ref{fig:Fe-X_LOOCV}) being the only failures.
Inspecting the matrix of available training data (Fig.~\ref{fig:systems}), it turns out that when the Fe-Sn or Fe-Sb system is quarantined, no other binary system containing Sn or Sb, respectively, remains in the training data. 
As a result, the model now has no training information about the behavior of systems containing these elements, and the predictions are pure, untrained guesses. 
In all other cases of quarantined Fe-$X$ systems, at least one other $X$-$Y$ system remains in the training data, thereby allowing the $X$ species to learn its behavior indirectly.

\subsection{Uncertainty quantification}
\label{sec:uncertainty_quantification}

\begin{figure}[h!]
    \centering
    \includegraphics[width=0.8\textwidth]{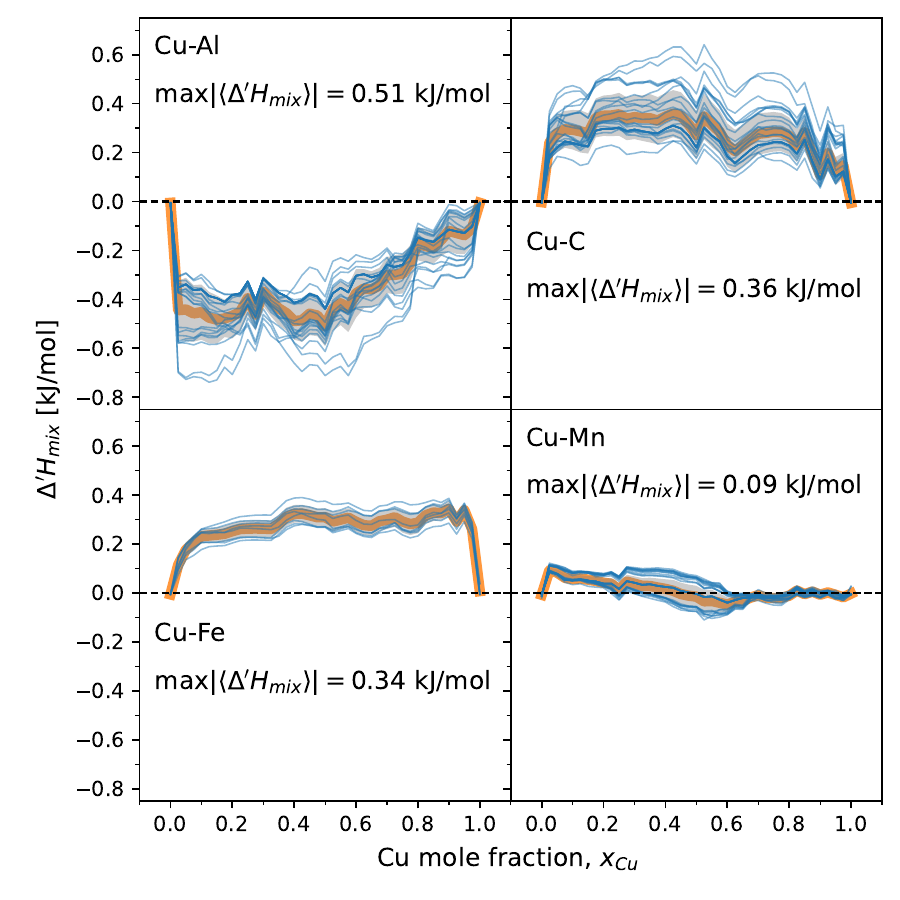}
    \caption{Example of uncertainty quantification for several Cu-$Y$ systems predicted using individually trained models with a second correlated binary system ($Y$-$Z$) removed (cf. Fig.~\ref{fig:systems}). 
    The $y$-axis represents the difference between the predicted mixing enthalpy ($H_{\text{mix}}$) and the one obtained from CALPHAD. 
    Further, the prime in $\Delta' H_{\text{mix}}$, indicates that two systems were quarantined, and the result is plotted with blue lines. 
    The thick orange line is the mean of the error of all individual models. 
    The gray error band represents the standard deviation band of our predictions.     
    }
    \label{fig_uncertainty}
\end{figure}

Since the model performs well overall ($k$-fold validation, Sec.~\ref{sec:k-fold} and Fig.~\ref{fig:parity_plot}) as well as in predicting the trends of $H_{\mathrm{mix}}$ for new binary systems (LOOCV procedure, Sec.~\ref{sec:LOOCV} and Fig.~\ref{fig:Fe-X_LOOCV}), we now attempt to quantify the uncertainty of the predicted $H_{\mathrm{mix}}$. To do this, we select a specific binary system $X$-$Y$ (e.g., Cu-Fe) for which we want to determine the uncertainty.
We call it the \textit{primary} system.
Subsequently, we selected another binary (\textit{secondary}) system that is correlated with the primary system, i.e., $Y$-$Z$ (e.g., Fe-Al).
The NN was then trained from scratch on a dataset in which both of these binary systems, primary and secondary, were quarantined.
Using this NN, $H_{\mathrm{mix}}$ of the primary system was predicted.
Then another secondary binary system that correlated with the primary binary system was removed, e.g., Fe-Cr, and a new NN was trained using all data except the primary system and the new secondary system.
A new prediction for the primary system was made using this NN. 
This process was repeated until all binary systems correlated with the primary system were covered. 
As a result, we obtained multiple predictions for the primary system (Cu-Fe in our example).

As shown in Fig.~\ref{fig_uncertainty}, removing certain binary systems from the training data did not significantly affect the prediction, while excluding other binary systems had a slightly greater impact on the accuracy of the prediction.
The Cu-Fe system shows the best overall predictability due to the fact that the training data contains numerous Fe-$Z$ systems. 
Thus, despite the quarantining of a second binary system, there are many other Fe-containing systems from which the NN can learn about binary interactions of Fe with other species.
Overall, all individual models with two quarantined systems yield consistent predictions, with the standard deviation $\lessapprox0.5\uu{kJ/mol}$, a value an order of magnitude smaller than the spread of experimental data (see green crosses in Fig.~\ref{fig:Fe-X_LOOCV}).

\subsection{Redlich-Kister parameter extraction}
\label{sec:redlich-kister_results}

\begin{figure}
    \centering
    \includegraphics[width=0.8\linewidth]{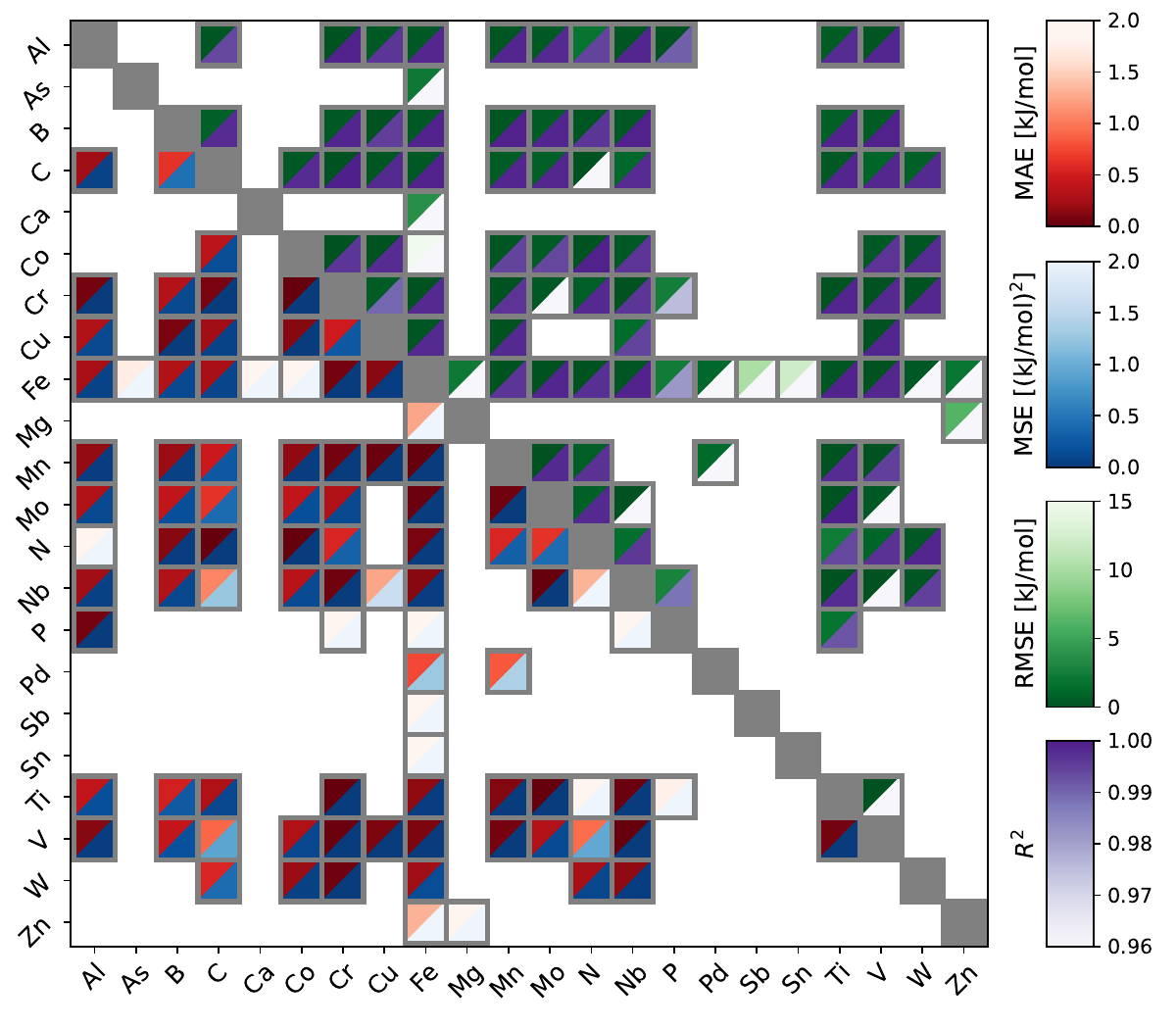}
    \caption{Performance evaluation of the final NN model (Fold 7) on the systems where input data (thermodynamic database~\cite{matcalc} and, if available, also experimental data; cf. Fig.~\ref{fig:systems}) is available, using four common metrics: MAE and MSE are visualized in the lower triangle, while RSME and $R^2$ are shown in the upper triangle.}
    \label{fig:metrics}
\end{figure}

The overall metrics of the final NN (Fold 7, Sec.~\ref{sec:k-fold}) are graphically summarized in Fig.~\ref{fig:metrics}.
The error metrics of Fold 7 seem to perform significantly well with the lowest mean absolute error of $16.52\uu{J/mol}$ and thus this fold was chosen to perform the RK parameter extraction. 
As demonstrated in Fig.~\ref{fig:metrics}, the mixing enthalpy predictions agree well with available CALPHAD, and thus the model was used to extrapolate to new binary systems that were not originally included in the dataset. 

We developed an in-house Python code to facilitate the extraction of the RK parameters, $L_k$, (Eq.~\eqref{eq:redlich-kister}) from the predicted $H_{\mathrm{mix}}$ data using our NN~\cite{data_share}.
This way, the predictions generated in this work can be readily reintegrated into thermodynamic databases.
The resulting RK parameters for all systems are tabulated in the Supplementary Material Table S3 and also in the project's online data share~\cite{data_share}.
An example of the final $L_k$ data for representative systems discussed in this paper, along with the values extracted from the used CALPHAD database, are given in Table~\ref{tab:rk_parameters}.

\begin{table}[h!]
    \centering
    \caption{Fitted Redlich-Kister (RK) parameters, $L_k$, for the binary systems Fe-Al, Fe-Cu, and Fe-Ti. 
    We provide them here with the same precision as typically coded in the thermodynamic database files while noting that this does not represent the physical accuracy of the fitted coefficients (see discussions in Secs.~\ref{sec:LOOCV} and \ref{sec:uncertainty_quantification}).}
    \label{tab:rk_parameters}
    \begin{tabular}{crrr}
        \toprule
        \textbf{Parameter} & \multicolumn{1}{c}{\textbf{Fe-Al}}  &  \multicolumn{1}{c}{\textbf{Fe-Cu}} & \multicolumn{1}{c}{\textbf{Fe-Ti}} \\
        & \multicolumn{1}{c}{[J/mol]} & \multicolumn{1}{c}{[J/mol]} & \multicolumn{1}{c}{[J/mol]} \\ 
        \midrule
        \( L_1 \) & $-75142.12$ &  $44798.18$ & $-71265.53$ \\ 
        \( L_2 \) & $9925.93$ & $-8590.68$ & $-7019.46$ \\ 
        \( L_3 \) & $-13731.09$ & $2916.35$ & $12248.17$ \\ 
        \( L_4 \) & $-11671.44$ & $672.81$ & $-592.52$ \\ \hline
        \bottomrule
    \end{tabular}
\end{table}

\begin{figure}[h!]
    \centering
    \includegraphics[width=0.8\textwidth]{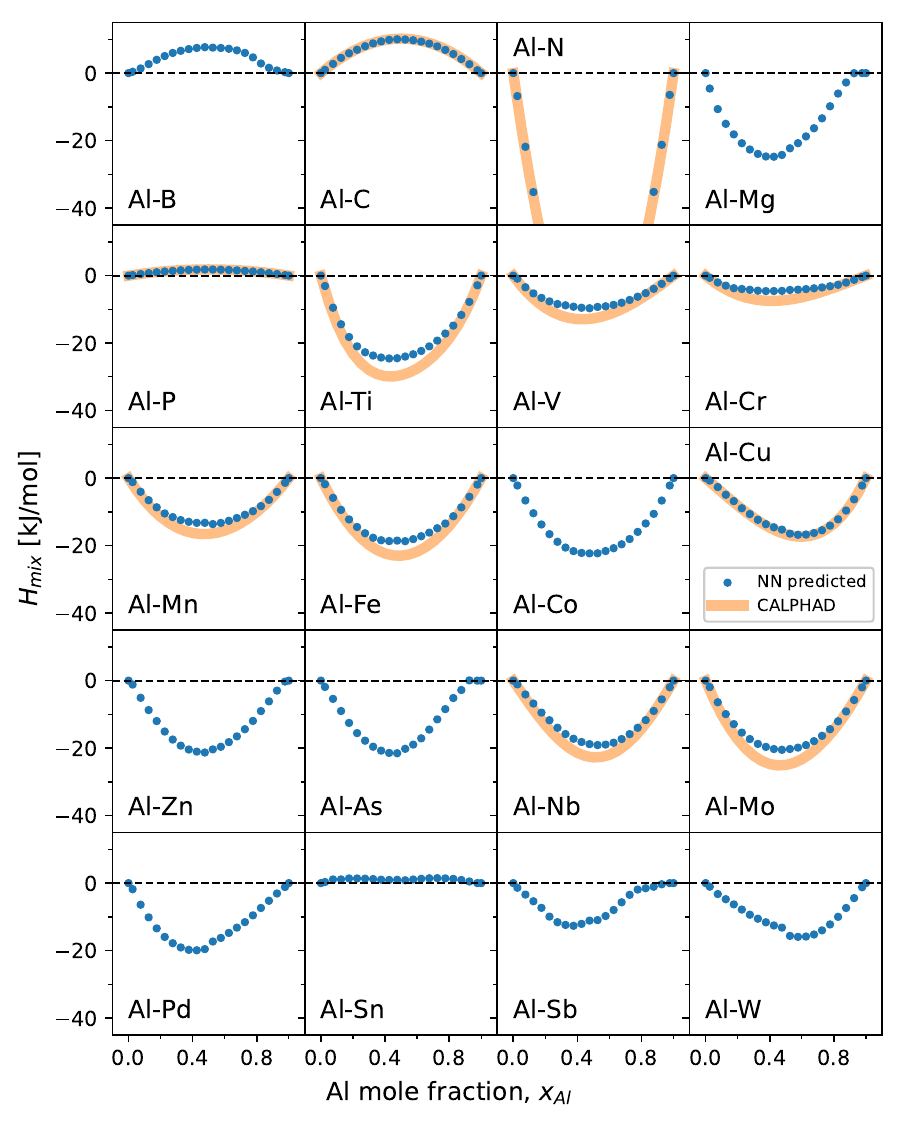}    
    \caption{Predicted mixing enthalpy ($H_{\mathrm{mix}}$) for all 20 Al-$X$ binary systems. 
    If available in the thermodynamic database, the CALPHAD results are shown with a solid orange line for reference.}
    \label{fig_all_Al_binary}
\end{figure}

Finally, in Fig.~\ref{fig_all_Al_binary}, we show an example of the predictions of the model for the Al-$X$ binary system.
When our training data also contained the respective system extracted from the open thermodynamic database~\cite{matcalc}, we also plotted them with a solid line.
The rest of the systems are pure predictions. 
A similar graph for the Fe-$X$ systems is shown in Supplementary Figure S1.
While the predictions are largely unchanged w.r.t. Fig.~\ref{fig:Fe-X_LOOCV}, the predictions for the Fe-Sn and Fe-Sb systems are now correct since some information about these species is now included in the training dataset.

These final predictions are done using a model that was trained using data from all binary systems (see Sec .~\ref {sec:k-fold}, Fold 7). 
Since the Fold 7 model performed well in the LOOCV evaluations (Sec.~\ref{sec:LOOCV}), where one binary system is quarantined and is used for predictions, it is assumed that the model understands the underlying correlations between the features and the targeted mixing enthalpy. 
However, it was also observed that certain binary systems with not many correlated elements in the training data (As-$X$, Sn-$X$, Sb-$X$) seem to perform relatively worse than systems that have an abundance of correlated elements. 
Consequently, this may also be the case for the extrapolated predictions of the new binary systems that were not included in the training data. 
Nonetheless, keeping in mind the results from the uncertainty quantification, where two related binary systems were removed and the error was analysed, we believe that even for the extrapolated predictions, the difference in error between the true mixing enthalpy and the predicted mixing enthalpy should be less than $1\uu{kJ/mol}$.

\section{Summary}
\label{sec:conclusions}

In this study, we successfully developed and validated a neural network model for predicting the mixing enthalpy of binary liquid phases using a thermodynamic database for steel, amended with experimental data from the literature. 
The input data covered 91 systems out of 231 possible for the selected 22 elements. 
The input data was characterized using elemental features together with compositionally dependent ones, which were extracted using Matminer.
Our neural network model was validated for its out-of-bounds predictive power using a leave-one-system-out cross-validation approach and an uncertainty quantification.
It achieved a mean absolute error (MAE) of $16.52\uu{J/mol}$ and an $R^2$ score of $0.99$, indicating a strong correlation between the predicted and actual values.

We also implemented an in-house Python code to extract the Redlich-Kister parameters from the predicted mixing enthalpy data.
The tabulated results, available in a tabulated form from the project's repository~\cite{data_share}, can be readily reintegrated into a thermodynamic database.
We note that the presented model works at a constant temperature ($1873\uu{K}$), as we aimed to prove the model's ability to predict for systems not seen during training.
The training dataset can be extended to include temperature-dependent data, which, however, is beyond the scope of this work.

We envision that this approach will pave the way for routinely employing ML to augment existing databases for missing data and, together with other ML algorithms (e.g.~\cite{Laiu2022, DEFFRENNES2024102745}), will enable predictions for higher-order systems, eventually forming a truly multi-component database. 
On a more general level, our study demonstrates the potential of neural network models in advancing materials informatics by providing accurate and reliable predictions of mixing enthalpy for binary systems.

\section*{Acknowledgments}
The financial support by the Austrian Federal Ministry for Digital and Economic Affairs, the National Foundation for Research, Technology and Development and the Christian Doppler Research Association is gratefully acknowledged. We would also like to express our deepest gratitude to Mat\v{e}j Holec for his work in collecting experimental data from the literature.

\section*{Data availability}
The raw data and materials supporting the conclusions of this article are available upon reasonable request to the corresponding author or can be directly downloaded from \cite{data_share}.

\bibliographystyle{elsarticle-num} 
\bibliography{Hmix_ML_paper_Clement}

\end{document}